\documentclass[aps,twocolumn,pra,longbibliography,superscriptaddress,10pt]{revtex4-2}

\usepackage[T1]{fontenc}
\usepackage{dsfont}
\usepackage{amsfonts}
\usepackage{amsmath}
\usepackage{bbold}
\usepackage{amssymb}
\usepackage{graphicx}
\usepackage{braket}
\usepackage{xcolor}
\usepackage{diagbox}
\usepackage{comment}
\usepackage{booktabs}

\usepackage[colorlinks=true,urlcolor=blue,citecolor=blue,linkcolor=blue]{hyperref}

\newcommand{\la}{\lambda}
\def \Tr {\text{Tr}}
\newcommand{\Hs}{\mathcal{H}}
\newtheorem{theorem}{Theorem}
\newtheorem{prop}{Proposition}
\newtheorem{conjecture}{Conjecture}

\newcommand{\HSs}{\mathcal{B}}

\newcommand*{\sbinom}[2]{\ensuremath{\Bigl(\genfrac{}{}{0pt}{}{#1}{#2}\Bigr)}}

\newcommand{\rema}[1]{\textcolor{purple}{#1}}

\begin{document} 

\widetext

\title{Nonequivalence between absolute separability and positive
partial transposition\\ in the symmetric subspace}

\author{Jonathan Louvet}
\email{jlouvet@uliege.be}
\affiliation{Institut de Physique Nucléaire, Atomique et de Spectroscopie, CESAM, University of Liège, B-4000 Liège, Belgium}

\author{Eduardo Serrano-Ens\'astiga}
\email{ed.ensastiga@uliege.be}
\affiliation{Institut de Physique Nucléaire, Atomique et de Spectroscopie, CESAM, University of Liège, B-4000 Liège, Belgium}

\author{Thierry Bastin}
\email{t.bastin@uliege.be}
\affiliation{Institut de Physique Nucléaire, Atomique et de Spectroscopie, CESAM, University of Liège, B-4000 Liège, Belgium}

\author{John Martin}
\email{jmartin@uliege.be}
\affiliation{Institut de Physique Nucléaire, Atomique et de Spectroscopie, CESAM, University of Liège, B-4000 Liège, Belgium}

\date{\today}

\begin{abstract}
The equivalence between absolutely separable states and absolutely positive partial transposed (PPT) states in
general remains an open problem in quantum entanglement theory. In this work, we study an analogous question
for symmetric multiqubit states. We show that symmetric absolutely PPT (SAPPT) states (symmetric states that
remain PPT after any symmetry-preserving unitary evolution) are not always symmetric absolutely separable by
providing explicit counterexamples. More precisely, we construct a family of entangled five-qubit SAPPT states.
Similar counterexamples for larger odd numbers of qubits are identified.
\end{abstract}
\maketitle

\section{Introduction} 
The study of the properties of quantum states subject to unitary evolutions has been at the heart of quantum mechanics since its foundation. One of the most intriguing properties of a multipartite quantum state is entanglement, which can be created or destroyed by unitary evolutions. In a region surrounding the maximally mixed state, there exists a set of mixed states, called \emph{absolutely separable} (AS), which remain separable under any global unitary transformation~\cite{PhysRevA.63.032307,PhysRevA.66.062311,PhysRevA.75.062330}. Characterizing this set is crucial as it defines the minimum conditions a quantum system must meet to attain a specific level of entanglement after a unitary transformation. The AS set in the bipartite scenario has been shown to be convex and compact ~\cite{PhysRevA.89.052304}, and its boundary has been determined for qubit-qudit systems ~\cite{PhysRevA.103.052431}. A complete characterization of the AS set is still lacking, however, and remains an important open problem in quantum information science~\cite{IQOQI}. Absolute separability criteria can always be expressed in terms of mixed state eigenvalues, as these are the only quantities that are invariant under unitary transformations. Because of this, AS states are also called \emph{separable from spectrum}.

The well-known positive partial transposition (PPT) criterion introduced by Peres~\cite{PhysRevLett.77.1413} is a simple and powerful method, and therefore very useful, for testing entanglement, although it only provides a sufficient and not a necessary condition. By this criterion, a multipartite quantum state $\rho$ is entangled in the $A|B$ bipartition if it is negative under partial transposition (NPT), i.e., $\rho^{T_A}$ has at least one negative eigenvalue.
This condition has been shown, in fact, to be necessary and sufficient for qubit-qubit and qubit-qutrit systems only~\cite{10.1007/BF02391860,WORONOWICZ1976165,HORODECKI19961}. This indicates the existence of entangled PPT states, also known as \emph{bound entangled states}. Many explicit bound entangled states have been presented and studied~\cite{PhysRevA.85.060302,PhysRevA.86.042316,brugues2016characterizing,PhysRevLett.80.5239}. In multipartite systems, the entanglement depends on the specific way the system is partitioned. A multipartite state is said to be (fully) separable if it can be written as a convex combination of multipartite product states~\cite{PhysRevA.61.042314}. Similarly, the PPT property depends on the bipartition considered. In this work, we are interested in states that are PPT for any bipartition, and we will call them just as PPT states.  \\ \indent
By analogy with the set of AS states, there is a set of absolutely PPT (APPT) states, i.e., states that remain PPT after any unitary evolution. In contrast with the AS set, the set of APPT states has been  fully characterized through linear matrix inequalities (LMIs)~\cite{PhysRevA.76.052325}. Thus, because of the close connection between separable states and PPT states, a promising way to solve the \textit{separability from spectrum} open problem would be to show that the APPT and AS sets are equivalent, as was undertaken in Ref.~\cite{arunachalam2015}. Although AS implies APPT according to the PPT criterion, the converse remains an open question.
So far, the equivalence of AS and APPT has only been proved for qubit-qudit systems~\cite{PhysRevA.88.062330} and for various families of states~\cite{arunachalam2015}. 
To disprove the general equivalence between the sets, it suffices to provide a counterexample, that is, an entangled APPT state. However, finding a counterexample is a remarkably difficult task, as many known strong entanglement criteria fail to detect entanglement in APPT states~\cite{arunachalam2015}. 
Other criteria for states to be APPT and AS have been explored for higher-dimensional systems~\cite{xiong2024:2408.11684,Abellanet-Vidal}, as well as the extremal points of the AS and APPT sets for qutrit-qudit systems~\cite{song2024}. \\ \indent The absolute separability problem arises in many forms, notably for systems with continuous variables \cite{PhysRevLett.90.047904,Lami_2018} and bosonic (symmetric) systems \cite{Abellanet-Vidal}. This work focuses on the latter case and more particularly on symmetric $N$-qubit systems. Their quantum states $\rho$ and evolution are restricted to the symmetric subspace of dimension $N+1$ within the much larger Hilbert space of dimension $2^N$. They satisfy $\rho = \pi \rho \pi'$ with $\pi$ and $\pi'$ any pair of permutation operators. The admissible unitary evolutions are also limited to $\mathrm{SU}(N+1)$ transformations. A symmetric state $\rho$ is said to be symmetric absolutely PPT (SAPPT) \cite{Abellanet-Vidal} if $U\rho U^{\dagger}$ is PPT for any $U \in \mathrm{SU}(N+1)$~\footnote{Another definition of SAPPT states for two qudits in a symmetric state has recently been given in Ref.~\cite{cha.joh.mac:21} as follows: A symmetric bipartite state $\rho \in \HSs(\Hs_d^{\vee 2})$ is SAPPT if $P (U \rho  U^{\dagger} )^{T_A} P \geq 0$ for any unitary transformation $U \in \HSs(\Hs_d^{\vee 2})$, where $P$ is the projector onto the subspace $\Hs_d^{\vee 2}$. However, this alternative definition is only defined for bipartite systems of balanced dimensions. Since we consider every possible bipartitions of multipartite states, which results in unbalanced bipartitions, this definition cannot be applied to our work.}. Similarly, a state is Symmetric Absolutely Separable (SAS) if the state remains separable after the action of any $U\in \mathrm{SU}(N+1)$. Due to the restriction on unitary operations that can be applied to the state, a SAS state is not necessarily AS~\cite{10.21468/SciPostPhys.15.3.120}. A full parametrization of SAS states has been given for $N=2$~\cite{10.21468/SciPostPhys.15.3.120}, and additional SAS and SAPPT witnesses have been constructed~\cite{10.21468/SciPostPhys.15.3.120,PhysRevA.109.022430,Filippov_2017,Abellanet-Vidal}. However, the SAS and SAPPT sets remain to be fully characterized. A question similar to that of the nonsymmetric case then arises: Are SAPPT states always SAS? This is true for symmetric two- and three-qubit systems, which boil down to qubit-qubit and qubit-qutrit systems for which the PPT criterion is necessary and sufficient. But what about larger qubit systems? This question represents much more than a simplified version of the nonsymmetric case; it is of significant independent interest because of the many physical quantum systems constrained by permutation symmetry, such as Bose-Einstein condensates~\cite{doi:10.1126/science.269.5221.198, PhysRevLett.132.173401,santiago2024spinentanglementantiferromagneticspin1} and multiphoton systems \cite{Ferretti:24}. In addition, it is of interest in the study of absolutely classical spin-$j$ states, which are equivalent to SAS states~\cite{PhysRevA.95.012318}. More generally, entanglement in bosonic systems plays an important role as a resource in quantum metrology and quantum information \cite{PhysRevX.10.041012}. In this work, we thus address the question of the general equivalence of SAPPT and SAS states. Our strategy is to search for counterexamples to the equivalence between the two sets. More specifically, we examine a subset of SAPPT states with a spectrum of a specific form and look for some entangled states in that subset. 

The paper is organized as follows. Section~\ref{Sec.2} presents the full parametrization of the SAPPT states across a uniparametric spectrum. We then prove in Sec.~\ref{Sec.3} that for an odd number of qubits, some of these SAPPT states are entangled. Finally, we provide concluding remarks in Sec.~\ref{Sec.4}. A summary table with all the acronyms used in this work, their meanings and definitions can be found in Appendix~\ref{App.Acro}.
\section{Uniparametric spectrum of SAPPT states}
\label{Sec.2}
The symmetric sector of the Hilbert space of a $N$-qubit system, $\Hs^{\vee N}$, with $\Hs$ the single-qubit state space, is spanned by the Dicke states
\begin{equation}
\label{Eq.Dicke}
    \ket{D^{(\alpha )}_N} = \binom{N}{\alpha}^{-1/2} \sum_{\sigma} P_{\sigma}\big| \underbrace{00 \dots 00 }_{N-\alpha}   \underbrace{11 \dots 11}_{\alpha} \big>
\end{equation}
for $\alpha = 0, \dots , N$, where the sum runs over all possible permutation operators $P_{\sigma}$ of the $N$ qubits. Let us now consider a symmetric state given by the mixture
\begin{equation}
\label{rho.Eq}
\begin{aligned}    
    \rho(p)  & = p\, \rho_0 + (1-p) \, \ket{\psi_0}\bra{\psi_0}
\end{aligned}
\end{equation}
of the maximally mixed state in the symmetric sector,
\begin{equation}
\rho_0=\frac{1}{N+1}\sum_{m=0}^{N} \ket{D^{(\alpha )}_N}\bra{D^{(\alpha )}_N}=\frac{\mathds{1}_{N+1}}{N+1},
\end{equation}
with probability $p$, and a pure symmetric state $\ket{\psi_0}\bra{\psi_0}$ with probability $1-p$, where $p \in [0,1]$. The spectrum of $\rho(p)$ consists of two distinct eigenvalues: one non-degenerate eigenvalue $1-\frac{Np}{N+1}$ and another $N$-fold degenerate eigenvalue $\frac{p}{N+1}$, i.e., $\left(1-\frac{Np}{N+1} , \frac{p}{N+1} , \dots , \frac{p}{N+1} \right)$. 

Now consider the bipartition of $k$ and $N-k$ qubits $A \big| B$. We will denote the bipartitions of qubit systems that are in a symmetric state by $N_A| N_B = k|N-k$ with $N_A$ the cardinality of the set $A$ (and the same for $B$) since, for such states, the explicit specification of which qubits belong to which partition is irrelevant. Without loss of generality, we assume $k\leq \lfloor N/2 \rfloor$. Since $\rho$ is a fully symmetric state, it has a support and an image in the tensor product of the symmetric sectors of subsystems $A$ and $B$. Therefore, $\rho$ is a convex combination of states in the bipartite Hilbert space $\Hs^{\vee k} \otimes \Hs^{\vee(N-k)}$. This Hilbert space with dimension $(k+1)\times(N-k+1)$ is spanned by the tensor products of the $k$- and $(N-k)$-qubit Dicke states, i.e.,~all the states $\ket{D^{(\alpha )}_{k}}\ket{D^{(\beta)}_{N-k}}\equiv \ket{D^{(\alpha )}_{k}}\otimes\ket{D^{(\beta)}_{N-k}}$ for $\alpha=0,\ldots,k$ and $\beta=0,\ldots,N-k$.

The partial transposed state $\rho^{T_A}$ of \eqref{rho.Eq} is the convex combination of $\rho_0^{T_A}$ and $\rho_{\psi_0}^{T_A} \equiv(\ket{\psi_0}\bra{\psi_0})^{T_A}$. 
The smallest eigenvalue of $\rho^{T_A}$, denoted by $\lambda_{\min}(\rho^{T_A})$, is lower bounded by
\begin{equation}
\begin{aligned}    
\label{Eq.Min.Eig}
&  \lambda_{\min} \left(\rho^{T_A} \right) \geq  \sigma \left(\ket{\psi_0},p \right)
\end{aligned}
\end{equation}
with 
\begin{equation}
\label{Eq.Def.sigma}
    \sigma (\ket{\psi_0},p) \equiv p\, \lambda_{\min} \left( \rho_0^{T_A} \right) + 
   (1-p)\, \lambda_{\min } \left( \rho_{\psi_0}^{T_A} \right) .
\end{equation}
This lower bound follows from the result that the minimum eigenvalue of a sum of two hermitian matrices, $A+B$, is greater than or equal to the sum of the minimum eigenvalues of each matrix $A$ and $B$~\cite{hor.joh:12}. In particular, the equality holds when $A$ and $B$ have a common eigenvector for their lowest eigenvalue.

Let us first focus on the density matrix $\rho_0^{T_A}$. After some algebra (see Appendix~\ref{App.Dicke}), we obtain that $\rho_0^{T_A}$ reads
\begin{equation}
\begin{aligned}
\label{Eq.rho0}
\rho_0^{T_A} ={}& \frac{1}{N+1}\left( \sum_{\alpha=0}^N \ket{D_N^{(\alpha )}}\bra{D_N^{(\alpha )}} \right)^{T_A}\\
    ={}& \frac{1}{N+1} \sum_{\alpha=0}^N \sum_{\beta,\gamma =0}^{\alpha} \chi(\alpha,\beta) \, \chi(\alpha,\gamma) \\
&
    \times 
\ket{D^{(\alpha-\gamma)}_{k}}\ket{D^{(\beta)}_{N-k}} 
\bra{D^{(\alpha-\beta)}_{k}}\bra{D^{(\gamma)}_{N-k}}   
, 
\end{aligned}
\end{equation}
where 
\begin{equation}
\label{Eq.chi}
\chi(\alpha,\beta) \equiv \left[ 
\frac{\sbinom{k}{\alpha-\beta} \sbinom{N-k}{\beta}}{\sbinom{N}{\alpha}}
\right]^{1/2} .
\end{equation}
We calculate in Appendix~\ref{App.C} the eigendecomposition of $\rho_0^{T_A}$. In particular, we prove that its minimum eigenvalue is given by 
\begin{equation}
\label{evrhoTA}
\begin{aligned}    
\lambda_{\min}( \rho_0^{T_A} ) = & \left[ (N+1)\tbinom{N}{k} \right]^{-1},
\end{aligned}
\end{equation}
and that the states $\ket{D^{(0)}_{k}}\ket{D^{(N-k)}_{N-k}}$ and $\ket{D^{(k)}_{k}}\ket{D^{(0)}_{N-k}}$ are particular eigenvectors of $\rho_{0}^{T_A}$ with this eigenvalue. 

Next, let us calculate $\lambda_{\min} ( \rho_{\psi_0}^{T_A} )$. We start with the Schmidt decomposition of $\ket{\psi_0}$ with respect to the bipartition $A\big| B$, 
\begin{equation}
    \ket{\psi_0} = \sum_{r=1}^{k+1} \sqrt{\Gamma_r} \ket{\phi^A_r} \ket{\phi^B_r} .
\end{equation}
with $\Gamma_r\geq 0$, $\sum_{r} \Gamma_r =1$ and, without loss of generality, we can assume that $\Gamma_r \geq \Gamma_{r+1}$. The set of states $\mathcal{A} = \{ \ket{\phi^A_r}\}_{r=1}^{k+1}$ and $\mathcal{B} = \{ \ket{\phi^B_r}\}_{r=1}^{k+1}$ form orthonormal bases of $\Hs^{\vee k}$ and $\Hs^{\vee(N-k)}$, respectively. The spectrum of $\rho_{\psi_0}^{T_A}$ can be calculated exactly~\cite{PhysRevA.65.032314,JOHNSTON20181}, with $\lambda_{\min} ( \rho_{\psi_0}^{T_A} ) = -\sqrt{\Gamma_1 \Gamma_2}$.

Using the previous results, we can now evaluate the lower bound $\sigma (\ket{\psi_0},p)$ in Eq.~\eqref{Eq.Min.Eig}, which gives
\begin{equation}
\sigma (\ket{\psi_0},p) =\frac{p}{(N+1)\binom{N}{k}} - (1-p)
 \sqrt{\Gamma_1 \Gamma_2} .
\end{equation}
We can observe that this bound depends only on the Schmidt coefficients of $\ket{\psi_0}$. To deduce when the state is SAPPT, we are now interested in minimizing $\sigma (\ket{\psi_0},p)$ on the unitary orbit of the state $\rho(p)$, i.e., on $\{U\rho(p) U^{\dagger}: U\in \mathrm{SU}(N+1)\}$. By Eq.~\eqref{rho.Eq},  
\begin{equation}
    U \rho(p) U^{\dagger} = p \rho_0 
    + (1-p)  \ket{\psi}\bra{\psi} ,
\end{equation}
where $\ket{\psi} = U\ket{\psi_0}$. This means that minimizing $\sigma (\ket{\psi_0},p)$ on the unitary orbit of \eqref{rho.Eq} reduces to minimizing $\sigma (\ket{\psi},p)$ on the states $\ket{\psi} \in \Hs^{\vee N}$. Thus,
\begin{equation}
\label{Eq.lambda.min}
\min_{U \in \mathrm{SU}(N+1)} \lambda_{\min} \left( U\rho(p)^{T_A} U^{\dagger} \right)
 \geq 
\min_{\ket{\psi} \in \Hs^{\vee N}}
\sigma \left( \ket{\psi},p \right) ,
\end{equation}
where the right-hand side is equal to
\begin{equation}
\label{Eq.Bound.Exp}
\frac{p}{(N+1)\binom{N}{k}} - \frac{1-p}{2} 
\end{equation}
and obtained for a state $\ket{\psi}$ with Schmidt coefficients $\Gamma_1 = \Gamma_2 =1/2$ and the rest equal to zero. 
Thus, $\min_{U \in \mathrm{SU}(N+1)} \lambda_{\min}(\rho(p)^{T_A}) \geq 0$, and consequently $\rho(p)$ is SAPPT on the bipartition $k\big| N-k$, whenever Eq.~\eqref{Eq.Bound.Exp} is greater or equal to zero. The strictest SAPPT condition is given by $k= \lfloor N/2 \rfloor$, where we can search for extremal values of $p$ by setting Eq.~\eqref{Eq.Bound.Exp} to zero. This leads to the sufficient condition of the following theorem, which describes a uniparametric set of SAPPT state spectra.

\begin{theorem}
\label{thm1}
    Any symmetric $N$-qubit state $\rho$ with a spectrum composed of $N+1$ non-zero eigenvalues of the form $\left(1- \frac{Np}{N+1}, \frac{p}{N+1},\dots ,\frac{p}{N+1} \right)$ is SAPPT if and only if   $p \in [ p_{\min} , 1 ]$ with
    \begin{equation*}
  p_{\min}=\frac{1}{1 +2 \left[ (N+1) \sbinom{N}{\lfloor N/2 \rfloor} \right]^{-1}} .
    \end{equation*}
\end{theorem}
The necessary condition to be SAPPT for this family of states is proven by finding an NPT state for any $p$ strictly smaller than $p_{\min}$. Consider the state $\rho(p)$ of Eq.~\eqref{rho.Eq} with $\ket{\psi_0}$ the $N$-qubit GHZ state $\ket{\text{GHZ}_N} \equiv (\ket{D_N^{(0)}} - \ket{D_N^{(N)}})/\sqrt{2}$. 
This state saturates the bound~\eqref{Eq.lambda.min}. Indeed,  $\rho(p)^{T_A}$ for the bipartition $k\big|N-k$ has an eigenvector $(\ket{D_{k}^{(0)}}\ket{D_{N-k}^{(N-k)}} + \ket{D_{k}^{(k)}}\ket{D_{N-k}^{(0)}})/\sqrt{2}$ with eigenvalue equal to Eq.~\eqref{Eq.Bound.Exp}, which is negative for $p < p_{\min}$, indicating that the state is NPT. We write in Table~\ref{tab:bounds} the value of $p_{\min}$ for several numbers of qubits. 

Following the same lines of reasoning, the sufficient condition
of Theorem \ref{thm1} can be generalized for
symmetric states $\ket{\psi} \in \Hs_d^{\vee N}$ of $N$-qudit systems, with $\Hs_d$ the Hilbert space of an individual qudit and $D\equiv\dim \Hs_d^{\vee N} = \binom{N+d-1}{d-1}$.
We observe by symbolic calculations that $\lambda_{\min}( \rho_0^{T_A} ) = \left[ D\tbinom{N}{k} \right]^{-1},$ for $d\leq 8$ and different values of $N\leq 15$. Since the strictest SAPPT condition is obtained again when $k = \lfloor N/2 \rfloor$, we can state the following conjecture.
\begin{conjecture}
\label{thm2}
Any symmetric $N$-qudit state $\rho$ with a spectrum composed of
     $D=\binom{N+d-1}{d-1} $ non-zero eigenvalues of the form $\left(1- \frac{(D-1)\,p}{D}, \frac{p}{D},\dots ,\frac{p}{D} \right)$, with $p \in [ p_{\min} , 1 ]$ where
    \begin{equation*}
  p_{\min}=\frac{1}{1 +2 \left[ D \sbinom{N}{\lfloor N/2 \rfloor} \right]^{-1}} ,
    \end{equation*}
is SAPPT.
\end{conjecture}

Theorem~\ref{thm2} is an improvement over a SAPPT criterion derived in Ref.~\cite{Abellanet-Vidal} using invertible linear maps of operators. 

\begin{table}[t]
    \centering
    \setlength{\tabcolsep}{15pt}
\begin{tabular}{c c c c}
\midrule
\midrule
   $N$ & $p_{\min} $ & $p_{\mathrm{ent}}^{W_N}$ & $p_{\mathrm{ent}}$
   \\[2pt] \midrule
    4 & $\frac{15}{16}$ & / & $\frac{15}{16}$
    \\[6pt]
    5 & $\frac{30}{31} \approx 0.96774$ & 0.96862 &  0.96953
    \\[6pt]
    6 & $\frac{70}{71}$ & / & $\frac{70}{71}$
    \\[6pt]
    7 & $\frac{140}{141} \approx 0.99291$ & 0.99302 & 0.99329
    \\[6pt]
    8 & $\frac{315}{316}$ & / & $\frac{315}{316}$
    \\[6pt]
    9 & $\frac{630}{631} \approx 0.99842$ & 0.99845 & 0.99849
    \\[6pt] 
    10 & $\frac{1386}{1387}$ & / & $\frac{1386}{1387}$ \\[2pt]
    \midrule \midrule
\end{tabular}
\caption{Particular values of the probability $p$ defining the state $\rho(p)$ given by Eq.~\eqref{rho.Eq} with $\ket{\psi_0}=\ket{\mathrm{GHZ}_N}$, which has an eigenspectrum $\left( 1- \frac{Np}{N+1}, \frac{p}{N+1},\dots ,\frac{p}{N+1} \right)$. First column: number of qubits. Second column: minimum value of $p$, $p_{\min}$ as it appears in Theorem~\ref{thm1}, for $\rho(p)$ to be SAPPT. Third column: maximum value $p^{W_N}_{\mathrm{ent}}$ such that the witnesses $W_N$ given in Eqs.~\eqref{W5}, \eqref{W7} and \eqref{W9} for $N=5,7$ and $9$, respectively, detect that $\rho(p)$ is entangled. Fourth column: value $p_{\mathrm{ent}}$ below which the state $\rho(p)$ is found to be entangled using the method described in Ref.~\cite{PhysRevA.96.032312}.}
    \label{tab:bounds}
\end{table}
\section{Entangled SAPPT states of qubits}
\label{Sec.3}
\begin{figure}[t]
    \centering
\includegraphics[width=0.99\linewidth]{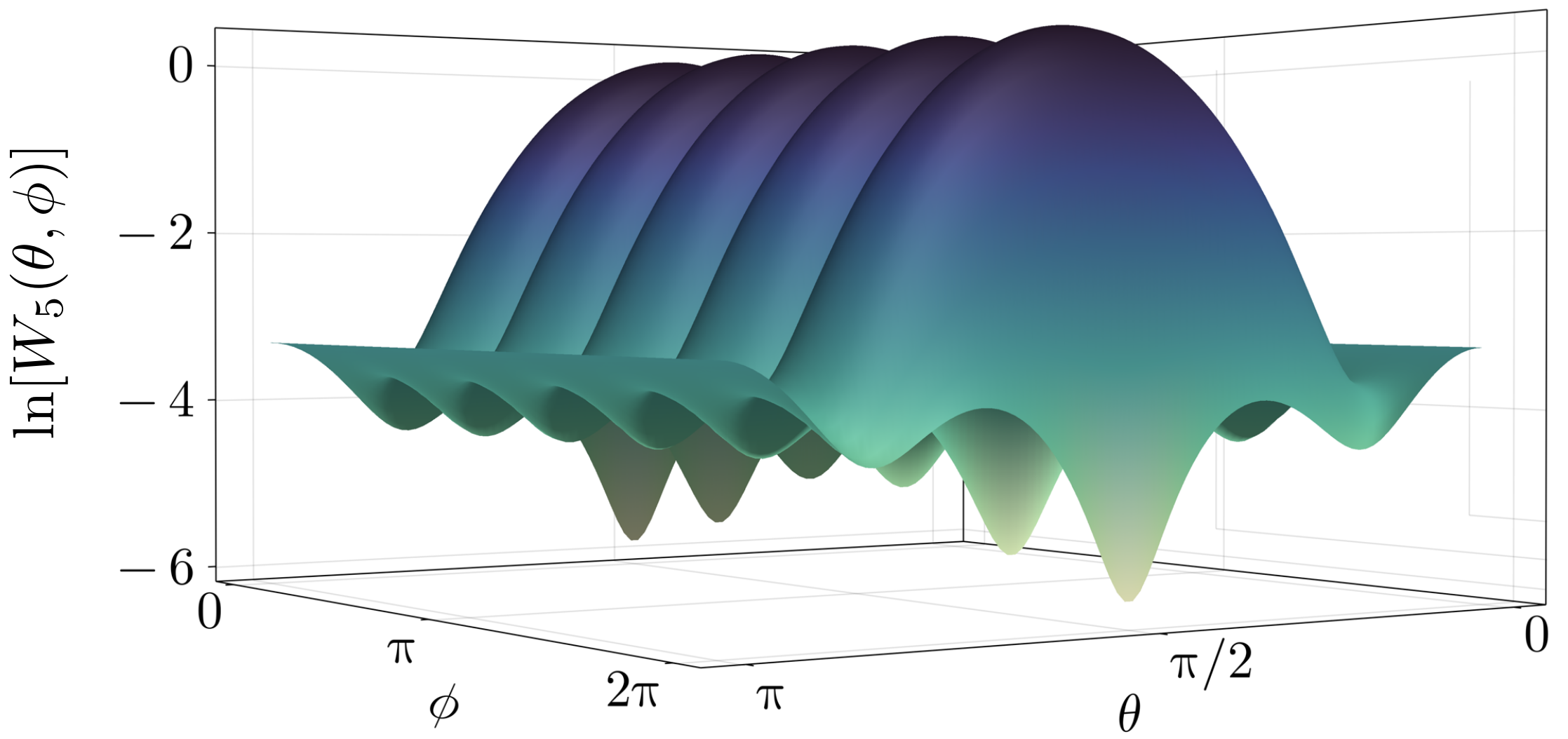}\\[10pt]
\includegraphics[width=0.97\linewidth]{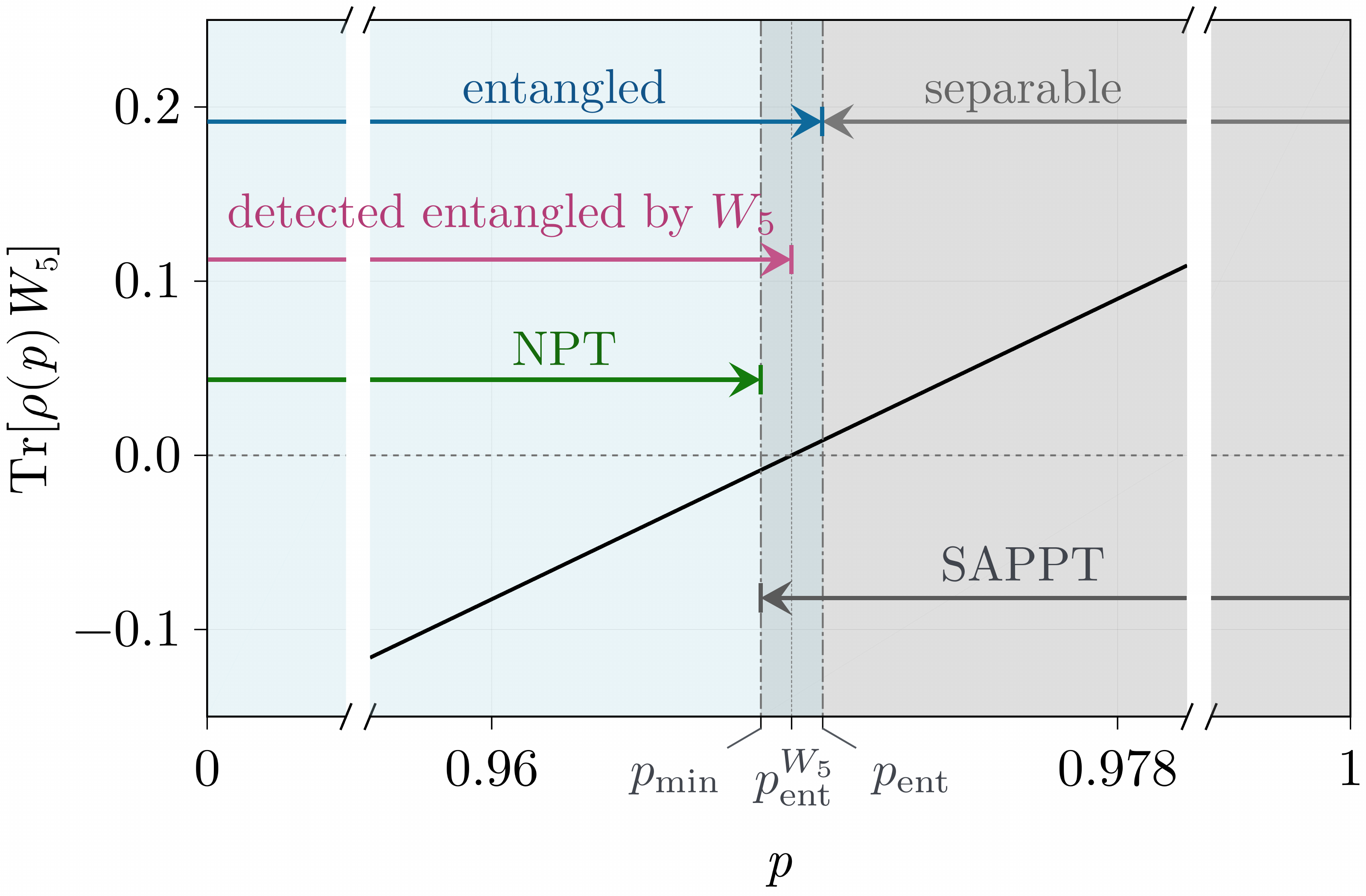}
    \caption{Top panel: Logarithm of the expectation value of the entanglement witness $W_5$ over the pure symmetric product states, showing the positivity of $W_5(\theta,\phi)$. Bottom panel: Expectation value of the entanglement witness $W_5$ in $\rho(p)$ as a function of $p$ (black oblique straight line). Entangled SAPPT states lie within the overlap between the blue (entangled states) and gray (SAPPT states) areas, ranging from $p=p_{\mathrm{min}}=30/31\approx 0.96774$ to $p_{\mathrm{ent}}= 0.96953$. Those detected by the witness $W_5$ given in Eq.~\eqref{W5} lie between $p_{\mathrm{min}}$ and $p_{\mathrm{ent}}^{W_5}\approx 0.96862<p_{\mathrm{ent}}$.}
    \label{fig:Wsep}
\end{figure}

We are now ready to present a family of SAPPT states that are not SAS, i.e., entangled SAPPT states. Consider now the state~\eqref{rho.Eq} for $5$ qubits with $\ket{\psi_0}=\ket{\mathrm{GHZ}_5}$. By Theorem~\ref{thm1}, $\rho(p)$ is SAPPT for $p \geq p_{\min}$ and NPT for $p< p_{\min}$. However, we find that the state $\rho(p_{\min})$ is detected to be entangled by not having a 2-copy PPT symmetric extension of the second party for the bipartition $1\big|4$ (see Ref.~\cite{Doherty2004} for more details). This can be checked using, for example, the QETLAB package~\cite{qetlab}. This method, whenever the state is entangled, gives additionally an entanglement witness~\cite{Doherty2004}. In our case, an entanglement witness for $\rho(p_{\min})$ is given in the Dicke-state basis of $\Hs^{\vee 5}$ by 
\begin{equation}
\label{W5}
W_5=
\begin{pmatrix}
a & 0 & 0 & 0 & 0 & c \\
0 & b & 0 & 0 & 0 & 0 \\
0 & 0 & 1 & 0 & 0 & 0 \\
0 & 0 & 0 & 1 & 0 & 0 \\
0 & 0 & 0 & 0 & b & 0 \\
c & 0 & 0 & 0 & 0 & a \\
\end{pmatrix} \;\mathrm{with}\;
\left\{
\begin{array}{l}
    a=0.0366656 \\
    b=-0.134595 \\
    c=-9.31947
\end{array}
\right. .
\end{equation}
One can easily check that indeed $\mathrm{Tr}(\rho(p_{\min}) W_5)\approx -0.0085 <0$. Moreover, one can check that $W_5$ is a proper entanglement witness for symmetric states by verifying that $W_5(\theta,\phi)\equiv \bra{\theta,\phi} W_5 \ket{\theta,\phi}\geq 0$ for all symmetric product states $\ket{\theta,\phi}=\ket{\varphi}^{\otimes N}$, with $\ket{\varphi}$ a single-qubit state parametrized as $\cos(\theta/2)\ket{0}+\sin(\theta/2)e^{i\phi}\ket{1}$. It suffices to check the positivity of the expectation value of $W_5$ on the set of pure product states $\ket{\theta, \phi}$ since any separable symmetric state $\rho_{\mathrm{sep}}$ can be written as a convex combination of them, i.e., $\rho_{\mathrm{sep}} = \int P(\theta,\phi) \ket{\theta,\phi} \bra{\theta,\phi} \mathrm{d} \Omega$ with $P(\theta,\phi)\geq 0$ and $\int P(\theta,\phi)\, \mathrm{d} \Omega = 1$~\cite{PhysRevLett.95.120502,PhysRevA.94.042343}. Thus, if the operator $W_5$ is such that $\bra{\theta,\phi} W_5 \ket{\theta,\phi} \geq 0$, then $\Tr (W_5 \rho_{\text{sep}}) \geq 0$ for all separable states. The minimal value of $W_5(\theta,\phi)$ is reached at $(\theta,\phi)=(\pi/2,0)$ and is approximately equal to $0.00276$. Figure~\ref{fig:Wsep} (top panel) shows the finite values of $\ln [W_5(\theta,\phi)]$, illustrating the positivity of $W_5(\theta,\phi)$. 

The operator $W_5$ given above also detects entanglement of $\rho(p)$ for values of $p$ other than $p_{\min}$ (see Figure~\ref{fig:Wsep}, bottom panel). In fact, $W_5$ detects that $\rho(p)$ is entangled for $ p  \leq p^{W_5}_{\mathrm{ent}} \approx 0.96862$, thus providing a uniparametric family of SAPPT bound entangled states $\rho(p)$ for $p \in [p_{\min} , p^{W_5}_{\mathrm{ent}}]$. A larger family can be obtained using the reformulation of the separability problem as a truncated moment problem (see Ref.~\cite{PhysRevA.96.032312} for more details), which can be implemented as a semidefinite optimization. Using this method, with a precision of $10^{-5}$ in the determination of the parameter $p$, the state $\rho(p)$ was found to be entangled for $ p_{\min} \leq p < p_{\mathrm{ent}} = 0.96953 $, and separable for $ p_{\mathrm{ent}} \leq p \leq  1 $. Finally, we should mention that there are separable SAPPT states that are not SAS. An example is given by $\rho(p_{\min})$ with $\ket{\psi_0}$ a symmetric product state $\ket{\theta,\phi}$.

In a similar way, we searched for entangled SAPPT states for a number of qubits up to $N=10$. For an odd number of qubits, the state~\eqref{rho.Eq} with $\ket{\psi_0}=\ket{\mathrm{GHZ}_N}$ is SAPPT and detected as entangled using QETLAB for values of $p$ in the range $[p_{\mathrm{min}}, p_{\mathrm{ent}}]$ (see Table~\ref{tab:bounds}). The respective entanglement witnesses for $N = 7,9$ are given in Appendix~\ref{App.EW}. On the other hand, for even $N$, we find that the state $\rho(p)$ is always separable for any $p \in [p_{\mathrm{min}}, 1]$. So we could not find an example of an entangled SAPPT state for an even number of qubits.

\section{Conclusions}
\label{Sec.4}
In this work, we established a sufficient condition for certain symmetric states of $N$-qudit systems to be SAPPT (symmetric absolutely PPT). For qubits, we showed that this condition is also necessary. In the course of this proof, we analytically determined the spectrum of $\rho_0^{T_A}$ where $\rho_0$ is the maximally mixed state in the symmetric subspace. Based on this, we proved the existence of entangled SAPPT states for an odd number of qubits from $N=5$ by constructing explicit entanglement witnesses. These results resolve an open question concerning the equivalence between SAPPT and SAS states, by showing that this equivalence does not hold in general, although it does apply to 2-qubit and 3-qubit systems. On the other hand, we have not yet identified entangled SAPPT states for an even number of qubits. It is important to note that in the non-symmetric case, the entangled SAPPT states presented in this work do not refute the possibility of the equivalence between APPT and AS states. If APPT states were indeed equivalent to AS states, the existence of entangled SAPPT states would provide a compelling example of how perfect indistinguishability among the constituents of multipartite systems, such as in bosonic systems, can profoundly affect their entanglement properties. We hope that our results will stimulate further exploration of this line of research in order to deepen our understanding of the interplay between symmetry and entanglement in quantum systems.
\section*{Acknowledgments} 
E.S.-E.\ acknowledges support from the postdoctoral fellowship of the IPD-STEMA program of the University of Liège (Belgium). J.M., T.B.\ and E.S.-E.\ acknowledge the FWO and the F.R.S.-FNRS for their funding as part of the Excellence of Science program (EOS project 40007526). T.B.\ also acknowledges
financial support through IISN convention 4.4512.08. Computational resources were provided by the Consortium des Equipements de Calcul Intensif (CECI), funded by the Fonds de la Recherche Scientifique de Belgique (F.R.S.-FNRS) under Grant No. 2.5020.11.
\appendix
\section{Acronyms}
\label{App.Acro}
Table~\ref{Table:2} contains all the acronyms used in this work.
\renewcommand{\arraystretch}{1.5} 
\begin{table*}[]
\begin{tabular}{c c c}
\midrule
\midrule
Acronym & Meaning                                          & Definition                                                                                         \\ \midrule
AS      & Absolutely separable                             & $U\rho U^{\dagger}$ is separable for all unitary $U$                                                \\ 
SAS     & Symmetric Absolutely Separable                   & $U_S\rho_S U_S^{\dagger}$ is separable for all symmetry preserving unitary $U_S$                \\ 
PPT     & Positive Partial Transposed  
& $\rho^{T_A}\geq 0$ 
\\ 
APPT    & Absolutely Positive Partial Transposed           & $(U\rho U^{\dagger})^{T_A}\geq 0$ for all unitary $U$                      
\\ 
SAPPT   & Symmetric Absolutely Positive Partial Transposed & $(U_S\rho_S U_S^{\dagger})^{T_A}\geq 0$ for all symmetry preserving unitary $U_S$ \\
\midrule
\midrule
\end{tabular}
\caption{Table of acronyms used in this work, their meaning and definition. The subscript $S$ refers to symmetric.}

\label{Table:2}
\end{table*}
\renewcommand{\arraystretch}{1} 
\section{Dicke states as bipartite states}
\label{App.Dicke}
For a given bipartition $k\big|N-k$ of $N$ qubits, the symmetric Dicke states $\ket{D_N^{(\alpha )}}$ defined in Eq.~\eqref{Eq.Dicke} can be rewritten as
\begin{equation}
    \begin{aligned}
    & \ket{D_N^{(\alpha )}} = 
\binom{N}{\alpha }^{-1/2} \sum_{r=0}^{\alpha} \left( \sum_{\sigma_{r}} P_{\sigma_{r}} \ket{\underbrace{ 0 0\dots 0  0}_{N-\alpha-r} \underbrace{1 1\dots 11}_{k+\alpha+r-N} } 
\right)
\\
&\phantom{\ket{D_N^{(\alpha )}} = \binom{N}{\alpha }^{1/2} } \otimes \left( \sum_{\lambda_{r}} P_{\lambda_{r}} \ket{\underbrace{ 00 \dots 00}_{r} \underbrace{11 \dots 11}_{N-k-r} } \right)
\\
& = \sum_{r=0}^{\alpha} \left[ 
    \frac{\binom{k}{N-\alpha-r} \binom{N-k}{r}}{\binom{N}{\alpha }}
    \right]^{1/2}  \ket{D_k^{(k+\alpha+r-N)}} 
\ket{D_{N-k}^{(N-k-r)}} ,
    \end{aligned}
\end{equation}
where $P_{\sigma_{r}}$ (respectively, ~$P_
{\lambda_r}$), are permutation operators of $k$ (respectively,~$N-k$) qubits. After the change of variable from $r$ to $\beta=N-k-r$, we get 
\begin{equation*}
\ket{D_N^{(\alpha )}} = \sum_{\beta=0}^{\alpha} \chi(\alpha,\beta) \ket{D_k^{(\alpha-\beta)}} \ket{D_{N-k}^{(\beta)}} , 
\end{equation*}
with 
\begin{equation*}
    \chi(\alpha,\beta) = 
    \left[ 
    \frac{\binom{k}{\alpha-\beta} \binom{N-k}{\beta}}{\binom{N}{\alpha}}
    \right]^{1/2}
    =
    \left[ 
    \frac{\binom{\alpha}{\beta}\binom{N-\alpha}{k+\beta-\alpha}}{\binom{N}{k}}
    \right]^{1/2} 
\end{equation*}
as defined in Eq.~\eqref{Eq.chi}. It is precisely this last equation that we use to write $\rho_0^{T_A}$ as in Eq.~\eqref{Eq.rho0}.
\section{Eigenspectrum of \texorpdfstring{$\rho_0^{T_A}$}{Lg}}
\label{App.C}
In this appendix, 
we calculate the eigenspectrum of $\rho_0^{T_A} \equiv \frac{1}{N+1} \mathbb{1}_{N+1}^{T_A}$, with $\mathbb{1}_{N+1}$ the identity in the symmetric subspace of the $N$-qubit system and $T_A$ the partial transposition performed on the first $k$ qubits. To this aim, we consider a  similar procedure as that used to calculate the spectra of angular momentum operators. First, we define the analogue of the angular momentum operators $J_\pm$, $J_z$ for the Dicke states, which we denote by $K_\pm$ and $K_0$. For all $\alpha=0,\ldots,N$, we set
\begin{equation}
\label{Eq.K.operators}
\begin{aligned}
    K_+ \ket{D_N^{(\alpha)}} \equiv &{}\sqrt{(N-\alpha)(\alpha+1)} \ket{D_N^{(\alpha+1)}} ,
    \\[5pt]
    K_- \ket{D_N^{(\alpha)}} \equiv &{}\sqrt{\alpha\,(N-\alpha+1)} \ket{D_N^{(\alpha-1)}} ,
    \\
    K_0 \ket{D_N^{(\alpha)}} \equiv &{}\left( \frac{N}{2}-\alpha \right) \ket{D_N^{(\alpha)}}.
\end{aligned}
\end{equation}
These operators verify the commutation relations $[K_+ , K_-]=-2K_0$ and $[K_{\pm} , K_0] = \pm K_{\pm}$. In addition, $K_{\pm}^\dagger=K_{\mp}$. We now define new operators acting on $\Hs^{\vee k}\otimes \Hs^{\vee(N-k)}$ as follows
\begin{equation}
\label{Eq.M.operators}
    \begin{aligned}
        M_{\pm} \equiv K_{\mp}^1 - K_{\pm}^2 , 
        \quad 
        M_0 \equiv K_0^1 - K_0^2 ,
    \end{aligned}
\end{equation}
where the superscript $1$ (respectively,\ $2$) refers to the subspace $\Hs^{\vee k}$ (respectively,\ $\Hs^{\vee(N-k)}$) on which the operator acts. By direct calculation, we obtain that
\begin{equation}
\label{Comm.rho0}
[M_0, M_{\pm} ]= \pm M_{\pm} , \quad
    [M_{\pm}, \rho_0^{T_A}] = 
    [M_0 , \rho_0^{T_A}] = 
    0 . 
\end{equation}
By construction, a possible orthonormal basis of eigenvectors of $M_0$ are the Dicke product states $\ket{D_{k}^{(\alpha)}}\ket{D_{N-k}^{(\beta)}}$ $(\alpha=0,\dots, k;\beta=0,\dots ,N-k)$ with eigenvalue $k +\beta-\alpha - N/2$. These eigenvectors share the same eigenvalue as long as the difference $\beta-\alpha$ is the same. Hence, the eigenvalues of $M_0$ can be written as $\mu_m=m- N/2$ with $m=0,\dots ,N$. 
A general eigenstate of $M_0$ with eigenvalue $\mu_m$ has the following general form
\begin{equation}
\label{Eq.General.form}
\ket{\phi_m} =
\sum_{r=\max (0,m+k-N)}^{\min (k,m)} d_r \ket{D_k^{(k-r)}} \ket{D_{N-k}^{(m-r)}} .
\end{equation}
The action of the ladder operators $M_{\pm}$ over these eigenstates fulfills the following proposition.
\begin{prop}
\label{Prop.1}
    The equation $M_- \ket{\phi_m}=0$, with $\ket{\phi_m}$ an eigenstate of $M_0$ with respect to $\mu_m$, admits a solution if and only if $m\leq k$, in which case it is unique (up to a normalization constant). Similarly,  the equation $M_+ \ket{\phi_{N-m}}=0$, with $\ket{\phi_{N-m}}$ an eigenstate of $M_0$ with respect to $\mu_{N-m}$, admits a solution if and only if $m\leq k$, in which case it is again unique up to a normalization constant.
\end{prop}
\indent \emph{Proof.} 
Following Eq.~\eqref{Eq.General.form} and denoting the bounds of the summation as $r_1=\max(0,m+k-N)$ and $r_2=\min(k,m)$, the action of $M_-$ on $\ket{\phi_m}$ gives
\begin{widetext}
    \begin{multline}
        M_-\ket{\phi_m} = d_{r_1} \sqrt{r_1(k-r_1+1)} \ket{D_k^{(k-r_1+1)}} \ket{D_{N-k}^{(m-r_1)}}- d_{r_2} \sqrt{(m-r_2)(N-k-m+r_2+1)}
\ket{D_k^{(k-r_2)}} \ket{D_{N-k}^{(m-r_2-1)}}\\
+\sum_{r=r_1}^{r_2-1}
     \Big( d_{r+1}
    \sqrt{(r+1)(k-r)}
    - d_r
    \sqrt{(m-r)(N-k-m+r+1)}
    \Big) \ket{D_k^{(k-r)}} \ket{D_{N-k}^{(m-r-1)}}
.
    \end{multline}
\end{widetext}
To fulfill $M_-\ket{\phi_m}=0$, the coefficients $d_r$ must obey
\begin{equation}
    \begin{aligned}
         d_{r_1} = {}&  0 , \quad \text{if } r_1\neq 0 ,
        \\
         d_{r_2} = {}& 0 , \quad \text{if } r_2\neq m ,
    \end{aligned}
\end{equation}
and the recursive relation, $\forall r = r_1, \dots, r_2-1$,
\begin{equation}
         d_{r+1} = \left[ 
        \frac{(m-r)(N-k-m+r+1)}{(r+1)(k-r)}
        \right]^{1/2} d_{r}.
\end{equation}
If $r_1 \neq 0 $ or $r_2 \neq m$, all $d_r$ must be zero by the recursive relation. This is the case when $m>k$. Conversely, if $m\leq k$, $r_1=0$, $r_2=m$, and the recursive relation yields, $\forall r = 1, \ldots, m$,
\begin{equation}
\label{Eq.Coef.Sol}
    d_r = \left[ 
    \frac{\sbinom{k-r}{m-r}\sbinom{N-k-m+r}{r}}{\sbinom{k}{m}}
    \right]^{1/2} d_0. 
\end{equation}
This defines a unique state $|\phi_m\rangle$ up to a normalization constant. Using an alternative version of Vandermonde convolution in combinatorics~\cite{Riordan1968Combinatorial},
\begin{equation}
\label{Vdmond}
\sbinom{\alpha+\beta}{\gamma} =   \sum_{\kappa=0}^{\gamma} \sbinom{\alpha-\kappa}{\gamma-\kappa} \sbinom{\beta+\kappa-1}{\kappa} ,
\end{equation}
we easily obtain that the state $|\phi_m\rangle$ gets normalized with
\begin{equation}
\label{d0}
d_0 = \sbinom{k}{m}^{1/2} \sbinom{N-m+1}{m}^{-1/2}.
\end{equation}
For $M_+$, a similar derivation holds. $\square$
\\
\indent
We now describe the common eigenbasis of $\rho_0^{T_A}$ and $M_0$ as a Theorem. 
\begin{theorem}
    \label{thm3}
    The operators $\rho_0^{T_A}$ and $M_0$, defined in Eqs.~\eqref{Eq.rho0} and~\eqref{Eq.M.operators}, form a complete set of commuting observables (CSCO) whose common eigenvectors generate a basis of the Hilbert space $\Hs^{\vee k}\otimes \Hs^{\vee N-k}$ with $k\leq N/2$. The normalized eigenvectors $\ket{n, m}$ are identified by two quantum numbers $n=0,\dots , k$ and $m= n, \dots\rema{,} N-n$, and satisfy the eigenvalue equations
    \begin{equation}
    \label{Eq.Eigen.Teo2a}
    \left\{\:
    \begin{aligned}
    \rho_0^{T_A}\ket{n, m} ={}& \lambda_{n} \ket{n, m},\\[5pt] 
        M_0\ket{n, m} ={}& \mu_m\ket{n, m} ,
    \end{aligned}\right.
    \end{equation}
with 
\begin{equation}
    \label{Eq.Eigen.Teo2}
    \lambda_{n} = \frac{1}{N+1}\frac{\sbinom{N+1}{n}}{\sbinom{N}{k}},\qquad\mu_m = m- \frac{N}{2}.
    \end{equation}
The eigenvectors $\ket{n,n}$ and $\ket{n,N-n}$ explicitly read
    \begin{align}        
    \label{Eq.State.n.n}
    \ket{n,n} ={} & \mathcal{N} \sum_{r=0}^n c_r^{k, n} \ket{D_k^{(k-r)}}\ket{D_{N-k}^{(n-r)}}   ,
    \\
    \ket{n,N-n} ={} & \mathcal{N} \sum_{r=0}^n c_r^{N-k,n} \ket{D_k^{(n-r)}} \ket{D_{N-k}^{(N-k-r)}} ,
    \end{align}
    with $\mathcal{N} = \sbinom{N-n+1}{n}^{-1/2}$ and
    \begin{equation}        
    \label{Eq.Coef.n.n}
    c_r^{k,n} = 
    \left[ 
    \sbinom{k-r}{n-r}\sbinom{N-k-n+r}{r}
    \right]^{1/2} .
\end{equation}    
For $n < m < N - n$, we have 
\begin{equation}
    \label{Eq.State.n.m}
    \begin{aligned}        
    \ket{n,m} \propto {}& 
    (M_+)^{m-n} \ket{n,n} ,
    \\
    \ket{n,N-m} \propto {}& 
    (M_-)^{m-n} \ket{n,N-n} 
    .
    \end{aligned}
\end{equation}
\end{theorem}
\emph{Proof} 
For each $n = 0, \ldots, k$, let us consider the unique (normalized) eigenstate $\ket{\phi_n}$ of $M_0$ with respect to $\mu_n$ that fulfills $M_- \ket{\phi_n}=0$. This state is given by Eqs.~\eqref{Eq.General.form}, \eqref{Eq.Coef.Sol}, and \eqref{d0} with $m = n$. The action of $\rho_0^{T_A}$ on it reads
\begin{multline}
    \label{Eq.First.rho0TA}
        (N+1)\rho_0^{T_A} \ket{\phi_n} = 
        \sum_{s,r=0}^n d_r\, \chi(k+n-r-s,n-s)
        \\
     \chi(k+n-r-s,n-r) \ket{D_k^{(k-s)}}
        \ket{D_{N-k}^{(n-s)}} .
\end{multline}
Using standard binomial identities~\cite{Riordan1968Combinatorial}, we get that
\begin{align}
\nonumber
  &  
  \chi(k+n-r-s,n-s)
  \,
  \chi(k+n-r-s,n-r)
\\[8pt]
\nonumber
={}& \frac{
\sbinom{k+n-r-s}{n-r}
\sbinom{N-k-n+r+s}{r}
}{
\sbinom{N}{k}
} \left[ \frac{
\sbinom{k-s}{n-s}
\sbinom{N-k-n+s}{s}
}{
\sbinom{k-r}{n-r}
\sbinom{N-k-n+r}{r}
}
\right]^{1/2} 
\\
={}& \frac{
\sbinom{k+n-r-s}{n-r}
\sbinom{N-k-n+r+s}{r}
}{
\sbinom{N}{k}
} \,\frac{d_s}{d_r} .
\end{align}
Thus,
\begin{align}
\label{Eq.Last.sum}
& \sum_{r=0}^n d_r\, \chi(k+n-r-s,n-s)
        \,\chi(k+n-r-s,n-r) \nonumber
\\
& =
\frac{d_s}{\sbinom{N}{k}} \sum_{r=0}^n \sbinom{k+n-r-s}{n-r}
\sbinom{N-k-n+r+s}{r} \nonumber
\\
& = 
\frac{\sbinom{N+1}{n}}{\sbinom{N}{k}} d_s ,
\end{align}
where we used Eq.~(\ref{Vdmond})
in the last equality. Inserting Eq.~\eqref{Eq.Last.sum} in Eq.~\eqref{Eq.First.rho0TA}, we obtain that $\rho_0^{T_A}\ket{\phi_n} = \lambda_n \ket{\phi_n}$. Hence the unique eigenstate $|\phi_n\rangle$ of $M_0$ with respect to $\mu_n$ that fulfills in addition $M_-\ket{\phi_n} = 0$ is also an eigenstate of $\rho_0^{T_A}$ with respect to $\lambda_n$. For $n = 0, \ldots, k$, we can denote accordingly these states by $\ket{\lambda_n,\mu_n}$ or just $\ket{n,n}$ for short.
They are given by Eqs.~\eqref{Eq.State.n.n} and \eqref{Eq.Coef.n.n}. In a very similar way, we obtain that the unique eigenstate $|\phi_{N-n}\rangle$ of $M_0$ with respect to $\mu_{N-n}$ that fulfills in addition $M_+\ket{\phi_{N-n}} = 0$ is also an eigenstate of $\rho_0^{T_A}$ with respect to $\lambda_n$. They can be denoted accordingly $\ket{n,N-n}$. \\
\indent 
Because of the commutation relations (\ref{Comm.rho0}), the state $M_+ \ket{n,n}$ is either zero or a common eigenstate of $\rho_0^{T_A}$ and $M_0$ with respective eigenvalues $\la_{n}$ and $\mu_{n+1}$. In the latter case, we can denote it by $\ket{n,n+1}$ after normalization. Again, $M_+ \ket{n,n+1}$ is either zero or generates a new common eigenstate $\ket{n,n+2}$. We can iterate this procedure a finite number of steps $q$ (since our Hilbert space is finite) until we get a state $\ket{n,n+q}$ that fulfills $M_+ \ket{n,n+q} =0$. In this case we know from above that this state must identify to $\ket{N-(n+q),n+q}$, and consequently $q = N - 2 n$. In this way, we obtain $N+1-2n$ common eigenstates $|n,m\rangle$ ($m = n, \ldots, N-n$) for each $n = 0,\dots , k$. The total number of these eigenstates amounts to
\begin{equation}
   \sum_{n=0}^k (N+1-2n)
    =
    (k+1)(N-k+1),
\end{equation}
which matches the dimension of the Hilbert space. 
\hfill $\square$
\begin{figure}
   \centering
    \includegraphics[width=0.85\linewidth]{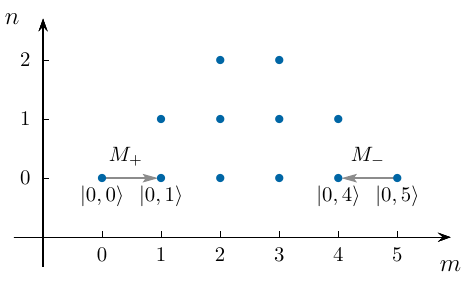}
    \caption{Schematic representation of the common eigenvectors $\ket{n,m}$ of $\rho_0^{T_A}$ and $M_0$ with respect to the quantum numbers $n$ and $m$ for $N=5$ and $k=2$. Each dot represents a common eigenvector and the dot ensemble 
    illustrates the allowed pairs of quantum numbers $(n,m)$.
    }
    \label{Table:3}
\end{figure}
\\
\indent
Let us give an example. For $N=5$ qubits and the bipartition of $k=2$ and $N-k=3$ qubits, we present in Fig.~\ref{Table:3} the allowed quantum numbers defining $(k+1)(N-k+1)=12$ eigenstates. We note that the eigenspectrum of $\rho_0^{T_A}$ is highly degenerate. The most-left (most-right) points are associated to the states that are eliminated by $M_-$ ($M_+$). 
\\
\indent
Finally we may note that $0 < \lambda_n < 2/(N+1)$, for all $k$ and $N$.
\section{Entanglement witnesses for \texorpdfstring{$N>5$}{Lg}}
\label{App.EW}
For $N=7$, the state $\rho$ given by~\eqref{rho.Eq} with $\ket{\psi_0}=\ket{\mathrm{GHZ}_7}$ for $p = p_{\mathrm{min}} = 140/141$ is detected to be entangled by not having 2-copy PPT symmetric extension of the second party for the bipartitions $1\big|6$. An entanglement witness in the Dicke basis is given by
\begin{equation}
\label{W7}
    W_7=
\begin{pmatrix}
a & 0 & 0 & 0 & 0 & 0 & 0 & d \\
0 & b & 0 & 0 & 0 & 0 & 0 & 0 \\
0 & 0 & c & 0 & 0 & 0 & 0 & 0 \\
0 & 0 & 0 & 1 & 0 & 0 & 0 & 0 \\
0 & 0 & 0 & 0 & 1 & 0 & 0 & 0 \\
0 & 0 & 0 & 0 & 0 & c & 0 & 0 \\
0 & 0 & 0 & 0 & 0 & 0 & b & 0 \\
d & 0 & 0 & 0 & 0 & 0 & 0 & a
\end{pmatrix}
\end{equation}
with 
\begin{equation}
       \left\{
\begin{array}{l}
    a = 0.00197514 \\
    b = 0.0643064 \\
    c = -0.189017 \\
    d = -31.2405
\end{array}
\right.
\end{equation}
from which we get $\mathrm{Tr}(\rho W_7)\approx -0.0038 <0$ and $\min W_7(\theta,\phi) \approx 0.001975 $ at $(\theta,\phi)=(0,0)$. Taking into account the results of Theorem~\ref{thm1}, $W_7$ detects that $\rho(p)$ is an entangled SAPPT state for $p_{\min} \leq p  \leq p^{W_7}_{\mathrm{ent}}\approx 0.9930$. 

Similarly, for $N=9$, the state $\rho$ given by~\eqref{rho.Eq} with $\ket{\psi_0}=\ket{\mathrm{GHZ}_9}$ for $ p_{\mathrm{min}} = 630/631 $ is detected to be entangled by not having a 2-copy PPT symmetric extension of the second party for the bipartitions $4\big|5$. An entanglement witness in the Dicke basis is given by
\begin{equation}
\label{W9}
    W_9 = 
\begin{pmatrix}
a & 0 & 0 & 0 & 0 & 0 & 0 & 0 & 0 & e \\
0 & b & 0 & 0 & 0 & 0 & 0 & 0 & 0 & 0 \\
0 & 0 & c & 0 & 0 & 0 & 0 & 0 & 0 & 0 \\
0 & 0 & 0 & d & 0 & 0& 0 & 0 & 0 & 0 \\
0 & 0 & 0 & 0 & 1 & 0 & 0 & 0 & 0 & 0 \\
0 & 0 & 0 & 0 & 0 & 1 & 0 & 0 & 0 & 0 \\
0 & 0 & 0 & 0 & 0 & 0 & d & 0 & 0 & 0 \\
0 & 0 & 0 & 0 & 0 & 0 & 0 & c & 0 & 0 \\
0 & 0 & 0 & 0 & 0 & 0 & 0 & 0 & b & 0 \\
e & 0 & 0 & 0 & 0 & 0 & 0 & 0 & 0 & a 
\end{pmatrix}
\end{equation}
with
\begin{equation*}
    \left\{
\begin{array}{l}
    a = 0.00235791 \\
    b = -0.013747 \\
    c = 0.0621661 \\
    d = -0.1636915 \\
    e = -114.305
\end{array}
\right. .
\end{equation*}
We have $\mathrm{Tr}(\rho W_9)\approx -0.004 <0$ and $\min W_9(\theta,\phi) \approx 0.0002234 $ at $(\theta,\phi)=(0.381,0)$. In the range  $p_{\mathrm{min}} \leq p  \leq 1$, i.e., when the state is SAPPT, $W_9$ detects that $\rho$ is entangled for $p_{\min} \leq p  \leq p^{W_9}_{\mathrm{ent}}\approx 0.99845$. 
\newline\newline\newline\newline\newline
\bibliography{Refs_APP.bib}
\end{document}